\documentclass[10pt,a4paper]{llncs}
\usepackage[utf8]{inputenc}
\usepackage{amsmath,amsfonts,amssymb}
\usepackage{ulem,mathtools}
\usepackage{graphicx}
\usepackage[ruled,vlined,linesnumbered]{algorithm2e}
\DontPrintSemicolon
\usepackage{xcolor}
\usepackage{standalone}
\usepackage{tikz}
\usepackage[toc,page]{appendix}
\usepackage{multirow}
\usepackage{hyperref}
\hypersetup{
	unicode=true,
	colorlinks=true,
	citecolor=blue!70!black,
	filecolor=black,
	linkcolor=red!70!black,
	urlcolor=blue,
	pdfstartview={FitH},
}

\newcommand{\Fp}{\mathbb{F}_p}
\newcommand{\Fq}{\mathbb{F}_{p^2}}
\newcommand{\fraka}{\mathfrak{a}}

\newcommand{\frakl}{\mathfrak{l}}
\DeclareMathOperator{\Cl}{Cl}
\DeclareMathOperator{\End}{End}
\newcommand{\ZZ}{\ensuremath{\mathbb{Z}}}
\DeclareMathOperator{\Set}{\mathcal{S}}
\DeclareMathOperator{\Prob}{Prob}

\title{Stronger and Faster Side-Channel Protections for CSIDH}
\author{Daniel Cervantes-V\'azquez\inst{1}, Mathilde Chenu\inst{2,3},
Jes\'us-Javier Chi-Dom\'inguez\inst{1}, Luca De Feo\inst{4}, Francisco
Rodr\'iguez-Henr\'iquez\inst{1}, Benjamin Smith\inst{3,2}}

\institute{CINVESTAV - Centro de Investigaci\`{o}n y de Estudios Avanzados del Instituto Polit\'{e}cnico Nacional,
	Mexico City, Mexico
	\and
	École polytechnique, Institut Polytechnique de Paris, Palaiseau, France
	\and 
	Inria, équipe-projet GRACE, Université Paris--Saclay, France
	\and 
	Universit\'{e} Paris Saclay – UVSQ,
	Versailles, France}
\date{}

\begin{document}
\maketitle

\begin{abstract}
  CSIDH is a recent quantum-resistant primitive based on
  the difficulty of finding isogeny paths between supersingular
  curves.
    Recently, two constant-time versions of CSIDH have been proposed:
    first by Meyer, Campos and Reith,
    and then by Onuki, Aikawa, Yamazaki and Takagi.
    While both 
  offer protection
  against timing attacks and simple power consumption analysis,
  they are vulnerable to more powerful attacks such as fault injections.
  In this work, we identify and repair two oversights 
    in these algorithms that compromised their constant-time character.
    By exploiting Edwards arithmetic and optimal addition chains,
    we produce
  the fastest constant-time version of CSIDH to date.
  We then consider the stronger attack scenario of fault injection,
  which is relevant for the security of CSIDH static keys in embedded hardware.
  We propose and evaluate a dummy-free CSIDH algorithm.
  While these CSIDH variants are slower,
  their performance is still within a small constant factor of
  less-protected variants.
  Finally, we discuss derandomized CSIDH algorithms.
\end{abstract}

\paragraph{Note:} A previous version of this article incorrectly
claimed that a test in Algorithm~\ref{alg:OAYT} leaks information
through the timing channel. We are grateful to Prof. Onuki for point
out our mistake.

\section{
    Introduction
}

Isogeny-based cryptography was introduced by Couveignes~\cite{Cou06}, 
who defined a key exchange protocol similar to
Diffie--Hellman 
based on the
action of an ideal class group on a set of ordinary elliptic
curves. %
Couveignes' protocol was independently rediscovered by Rostovtsev and
Stolbunov~\cite{RS06,Stol}, who were the first to recognize its
potential as a post-quantum candidate.
Recent efforts to make this system practical
have put it back at the forefront
of research in post-quantum cryptography~\cite{DFKS18}.
A major breakthrough was achieved by Castryck, Lange,
Martindale, Panny, and Renes with CSIDH~\cite{CLMPR18}, a
reinterpretation of Couveignes' system using supersingular curves
defined over a prime field.

The first implementation of CSIDH completed a key exchange in
less than 0.1 seconds, and its performance has been further improved
by Meyer and Reith~\cite{MR18}. %
However, both~\cite{CLMPR18} and~\cite{MR18} recognized the difficulty
of implementing CSIDH with constant-time algorithms,
that is,
algorithms whose running time, sequence of operations,
and memory access patterns
do not depend on secret data.
The implementations of~\cite{CLMPR18} and~\cite{MR18}
are thus vulnerable to simple timing attacks. %

The first attempt at implementing CSIDH in constant-time was realized by
Bernstein, Lange, Martindale, and Panny~\cite{BLMP18}, but their
goal was to obtain a fully deterministic reversible circuit 
implementing the class group action, to be used in quantum
cryptanalyses. %
The distinct problem of efficient CSIDH implementation
with side-channel protection was first tackled by Jalali,
Azarderakhsh, Mozaffari Kermani, and Jao~\cite{cryptoeprint:2019:297},
and independently by Meyer, Campos, and Reith~\cite{MCR18},
whose work was improved by
Onuki, Aikawa, Yamazaki, and Takagi~\cite{OAYT19}.

The approach of Jalali \emph{et al.} is similar to that of~\cite{BLMP18},
in that they achieve a stronger notion of constant time (running time
independent from \emph{all} inputs), at the cost of allowing the
algorithm to fail with a small probability. %
In order to make the failure probability sufficiently low, they
introduce a large number of useless operations, which make the
performance significantly worse than the original CSIDH algorithm. %
This poor performance and possibility of failure
reduces the interest of this implementation; we will
not analyze it further here.

Meyer \emph{et al.} take a different path: 
the running time of their algorithm is
independent of the secret key, but not of the output
of an internal random number generator.
They claim a speed only $3.10$ times slower than the
unprotected algorithm in~\cite{MR18}.
Onuki \emph{et al.} introduced new improvements,
claiming a speed-up of $27.35 \%$
over Meyer \emph{et al.}, i.e., a net slow-down factor of $2.25$
compared to~\cite{MR18}.

\paragraph{Our contribution.}
In this work we take a new look at side-channel protected
implementations of CSIDH. %
We start by reviewing the implementations in~\cite{MCR18}
and~\cite{OAYT19}. %
We highlight a flaw that makes their constant-time claims
disputable, and propose a fix for it. %
Since this fix introduces some minor slow-downs, we report on the
performance of the revised algorithms.

Then, we introduce new optimizations to make both~\cite{MCR18}
and~\cite{OAYT19} faster: we improve isogeny formulas for the 
model, and we introduce the use of optimal addition chains in the
scalar multiplications. %
With these improvements, we obtain a version of CSIDH protected
against timing and some simple power analysis (SPA) attacks that is 39\% more efficient
than~\cite{MCR18}. 

Then, we shift our focus to stronger security models. %
All constant-time versions of CSIDH presented so far use so-called
``dummy operations'', i.e., computations whose result is not used,
but whose role is to hide
the conditional structure of the algorithm from timing and SPA attacks
that read the sequence of operations performed from a single power trace.
However, this countermeasure is easily defeated by fault-injection attacks,
where the adversary may modify values during the computation. %
We propose a new constant-time variant of CSIDH without dummy
operations as a first-line defence.
The new version is only twice as slow as the simple constant-time
version.

We conclude with a discussion of derandomized variants of CSIDH. %
The versions discussed previously are
``constant-time'' in the sense that their running time is uncorrelated
to the secret key, however it depends on some (necessarily secret) seed to
a PRNG. %
While this notion of ``constant-time'' is usually considered good
enough for side-channel protection, one may object that a compromise
of the PRNG or the seed generation would put the security of the
implementation at risk, even if the secret was securely generated
beforehand (with an uncomprised PRNG) as part of a long-term or static keypair. %
We observe that this dependence on additional randomness is not necessary:
a simple modification of CSIDH, already considered in isogeny-based signature
schemes~\cite{SeaSign,DPV19}, can easily be made constant-time and
free of randomness. %
Unfortunately this modification requires increasing substantially the
size of the base field, and is thus considerably slower and not
compatible with the original version. %
On the positive side, the increased field size makes it much more
resistant to quantum attacks, a non-negligible asset in a context
where the quantum security of CSIDH is still unclear; it can thus be
seen as CSIDH variant for the paranoid.


\paragraph{Organization.}
In~\S\ref{sec:CSIDH} we briefly recall ideas, algorithms and
parameters from CSIDH~\cite{CLMPR18}. %
In~\S\ref{sec:repair} we highlight a shortcoming in~\cite{MCR18}
and~\cite{OAYT19} and propose a way to fix it. %
In~\S\ref{sec:optimize} we introduce new optimizations
compatible with all previous versions of CSIDH.
In~\S\ref{sec:fault} we introduce a new algorithm for evaluating
the CSIDH group action that is resistant against timing and some simple
power analysis attacks,
while providing protection against some fault injections. %
Finally, in~\S\ref{sec:derandomized} we discuss a more costly
variant of CSIDH with stronger security guarantees.

\paragraph{Notation.}
\textbf{M}, \textbf{S}, and \textbf{A}
denote the cost of computing a single multiplication,
squaring, and addition (or subtraction) in \(\Fp\),
respectively.
We assume that a constant-time equality test
\(\texttt{isequal}(X,Y)\) is defined,
returning \(1\) if \(X = Y\) and \(0\) otherwise.
We also assume that a constant-time conditional swap
\(\texttt{cswap}(X,Y,b)\) is defined,
exchanging \((X,Y)\) if \(b = 1\) (and not if \(b = 0\)).

\section{
    CSIDH
}
\label{sec:CSIDH}

CSIDH is an isogeny based primitive, similar to Diffie--Hellman, that
can be used for key exchange and encapsulation~\cite{CLMPR18},
signatures~\cite{SeaSign,DPV19,BKV19}, and other more advanced protocols. %
Compared to the other main isogeny-based primitive
SIDH~\cite{JDF11,DFJP14}, CSIDH is slower.
On the positive side, CSIDH has smaller public keys, 
is based on a better understood security assumption,
and supports an easy key validation procedure, making it
better suited than SIDH for CCA-secure encryption, static-dynamic and
static-static key exchange. %
In this work we will use the jargon of key exchange when we refer to
cryptographic concepts.

CSIDH works over a finite field $\Fp$, where $p$ is a prime of the
special form 
\[
  p := 4\prod_{i=1}^n\ell_i - 1
\]
with $\ell_1,\dots,\ell_n$ a set of small odd primes. %
Concretely, the original CSIDH article~\cite{CLMPR18} 
defined a 511-bit~$p$ 
with $\ell_1,\dots,\ell_{n-1}$ the first 73 odd primes,
and $\ell_n=587$.

The set of public keys in CSIDH is a subset of all supersingular
elliptic curves defined over $\Fp$,
in \emph{Montgomery form}
$y^2=x^3+Ax^2+x$, where $A\in\Fp$ is called the \emph{$A$-coefficient}
of the curve.\footnote{%
    Following~\cite{CLN16}, we represent \(A = A'/C'\) as a projective
    point \((A':C')\);
    see \S\ref{sec:montg}.
}
The endomorphism rings of these curves are isomorphic to orders in
the imaginary quadratic field $\mathbb{Q}(\sqrt{-4p})$. %
Castryck \emph{et al.}~\cite{CLMPR18} choose to
restrict the public keys to the \emph{horizontal isogeny class} of the
curve with $A=0$, so that all endomorphism rings are isomorphic to
$\mathbb{Z}[\sqrt{-p}]$.

\subsection{The class group action}

Let \(E/\Fp\) be an elliptic curve with \(\End(E) \cong \ZZ[\sqrt{-p}]\).
If \(\fraka\) is a nonzero ideal in \(\ZZ[\sqrt{-p}]\),
then it defines a finite subgroup 
\(E[\fraka] = \bigcap_{\alpha\in\fraka}\ker(\alpha)\),
where we identify each \(\alpha\) with its image in \(\End(E)\).
We then have a quotient isogeny
\(\phi: E \to E' = E/E[\fraka]\) with kernel \(\fraka\);
this isogeny and its codomain is well-defined up to isomorphism.
If \(\fraka = (\alpha)\) is principal,
then \(\phi \cong \alpha\) and \(E/E[\fraka] \cong E\).
Hence, we get an action of the ideal class group \(\Cl(\ZZ[\sqrt{-p}])\)
on the set of isomorphism classes of elliptic curves \(E\) 
over \(\Fp\)
with \(\End(E) \cong \ZZ[\sqrt{-p}]\);
this action is faithful and transitive.
We write \(\fraka\ast E\) for the image of (the class of) \(E\)
under the action of \(\fraka\), which is (the class of) \(E/E[\fraka]\) above.

For CSIDH, we are interested in computing the action of
small prime ideals.
Consider one of the primes \(\ell_i\) dividing \(p+1\);
the principal ideal \((\ell_i) \subset \ZZ[\sqrt{-p}]\)
splits into two primes,
namely
\(\frakl_i = (\ell_i,\pi-1)\) and
\(\bar\frakl_i = (\ell_i,\pi+1)\),
where $\pi$ is the element of $\mathbb{Z}[\sqrt{-p}]$ mapping to the Frobenius
endomorphism of the curves.
Since \(\bar\frakl_i\frakl_i = (\ell_i)\) is principal,
we have \(\bar\frakl_i = \frakl_i^{-1}\) in \(\Cl(\ZZ[\sqrt{-p}])\),
and hence
\[
    \bar\frakl_i\ast(\frakl_i\ast E)
    =
    \frakl_i\ast (\bar\frakl_i\ast E)
    =
    E
\]
for all \(E/\Fp\) with \(\End(E) \cong \ZZ[\sqrt{-p}]\).

\subsection{The CSIDH algorithm}\label{sec:csidh}

At the heart of CSIDH is an algorithm that evaluates the class group
action described above on any supersingular curve over \(\Fp\). %
Cryptographically, this plays the same role as modular exponentiation
in classic Diffie--Hellman. %

The input to the algorithm is an elliptic curve $E:y^2=x^3+Ax^2+x$,
represented by its $A$-coefficient, and an ideal class
$\mathfrak{a} = \prod_{i =1}^{n} \mathfrak{l}_i^{e_i},$ represented by
its list of exponents $(e_i,\dots,e_n)\in\mathbb{Z}^n$. %
The output is the ($A$-coefficient of the) elliptic curve
$\fraka * E = \frakl_1^{e_1} * \cdots * \frakl_n^{e_n} * E$.

The isogenies corresponding to \(\frakl_i=(\ell_i,\pi-1)\) can be
efficiently computed using V\'elu's formul\ae{} and their
generalizations: exploiting the fact that $\#E(\Fp)=p+1=4\prod\ell_i$,
one looks for a point $R$ of order $\ell_i$ in $E(\Fp)$ (i.e., a point
that is in the kernels of both the multiplication-by-$\ell_i$ map and
$\pi-1$), computes the isogeny $\phi:E\to E/\langle R\rangle$ with
kernel $\langle R\rangle$, and sets $\frakl_i*E=E/\langle R\rangle$. %
Iterating this procedure lets us compute $\frakl_i^e*E$ for any
exponent $e\ge 0$.

The isogenies corresponding to $\frakl_i^{-1}$ are computed in a
similar fashion: this time one looks for a point $R$ of order $\ell_i$
in the kernel of $\pi+1$, i.e., a point in $E(\Fq)$ of the form
\((x,iy)\) where both \(x\) and \(y\) are in \(\Fp\)
(since \(i = \sqrt{-1}\) is in \(\Fq\setminus\Fp\) and satisfies \(i^p = -i\)).
Then one proceeds as before, setting
$\frakl_i^{-1}*E=E/\langle R\rangle$.

In the sequel we assume that we are given an algorithm
\texttt{QuotientIsogeny} which, given a curve \(E/\Fp\) 
and a point \(R\) in \(E(\Fq)\), computes the quotient isogeny
\(\phi: E \to E' \cong E/\langle R\rangle\), and returns the pair
\((\phi,E')\). %
We refer to this operation as \emph{isogeny computation}.
Algorithm~\ref{CSIDH_orig}, taken from the original CSIDH
article~\cite{CLMPR18},
computes the class group action.

\begin{algorithm}[ht]
	\caption{The original CSIDH class group action algorithm
        for supersingular curves over $\Fp$ 
        where $p = 4\prod_{i=1}^n\ell_i - 1$.
        The choice of ideals 
        $\frakl_i = (\ell_i, \pi - 1)$, where \(\pi\) is the element of
        \(\mathbb{Q}(\sqrt{-p})\) is mapped to the $p$-th power Frobenius
        endomorphism on each curve in the isogeny class, is a system parameter. This algorithm constructs exactly \(|e_i|\) isogenies for each ideal~\(\frakl_i\).
    }
    \label{CSIDH_orig}
    \KwIn{$A \in \Fp$ such that \(E_A \colon y^2 = x^3 + Ax^2 + x\) is supersingular, 
    and an integer exponent vector $(e_1, \dots, e_n)$}
    \KwOut{$B$ such that $E_B: y^2 = x^3 + Bx^2 + x$ is $\frakl_1^{e_1}
    \ast \cdots \ast \frakl_n^{e_n} \ast E_A$,
    }
    %
    \(B \gets A\)
    \;
    \While{some $e_i \ne 0$}{
        Sample a random $x \in \Fp$
        \;
        $s \gets +1$ if $x^3 + Bx^2 + x$ is square in $\Fp$, else $s \gets -1$
        \;
        $S \gets \{i \mid e_i \ne 0, \text{sign}(e_i) = s\}$
        \;
        \If{$S \ne \emptyset$}{
	        $k \gets \prod_{i \in S} \ell_i$
	        \;
	        $Q \gets [(p+1)/k]P$, where $P$ is the projective point
	        with $x$-coordinate $x$.
	        \;
	        \For{$i \in S$}{
	            $R \gets [k/ \ell_i]Q$
	            \tcp*{Point to be used as kernel generator}
	            \If{$R \ne \infty$}{
	                $(E_B, \phi) \gets \texttt{QuotientIsogeny}(E_B,R)$
	                \;
	                $Q\gets \phi(Q)$
	                \;
	                $(k,e_i) \gets (k/\ell_i,e_i - s)$
	            }
	        }
        }
    }
    \Return{$B$}
\end{algorithm}

For cryptographic purposes, the exponent vectors
$(e_1,\dots,e_n)$ must be taken from a space of size at least
$2^{2\lambda}$, where $\lambda$ is the (classical) security
parameter. %
The CSIDH-512 parameters in~\cite{CLMPR18} take $n=74$,
and all $e_i$ in the interval $[-5, 5]$, so that
$74 \log_2(2\cdot 5 + 1) \simeq 255.99$, consistent with the NIST-1
security level.
With this choice, the implementation of~\cite{CLMPR18} 
computes one class group action in 40 ms on average. %
Meyer and Reith~\cite{MR18} further improved this to 36 ms on average.
Neither implementation is constant-time.

\subsection{The Meyer--Campos--Reith constant-time algorithm}

As Meyer, Campos and Reith observe in~\cite{MCR18},
Algorithm~\ref{CSIDH_orig}
performs fewer scalar multiplications 
when the key has the same number of positive and negative exponents 
than it does in the unbalanced case where these numbers differ.
Algorithm~\ref{CSIDH_orig} thus leaks information about the distribution of positive and negative exponents under timing attacks.  
Besides this, analysis of power traces would reveal the cost of each isogeny computation,
and the number of such isogenies computed, which would leak the exact exponents of the private key.

In view of this vulnerability,
Meyer, Campos and Reith proposed in \cite{MCR18} a constant-time
CSIDH algorithm whose running time does not depend on the private key
(though, unlike~\cite{cryptoeprint:2019:297}, it still varies due to randomness). 
The essential differences between the algorithm of~\cite{MCR18} 
and classic CSIDH are as follows.
First, to address the vulnerability to timing attacks, 
they choose to use only positive
exponents in $[0, 10]$ for each $\ell_i$, instead of $[-5, 5]$ in the
original version, while keeping the same prime $p = \prod_{i= 1}^{74} \ell_i -1$.
To mitigate power consumption analysis attacks, their algorithm
always computes the maximal amount of isogenies allowed by the
exponent, using dummy isogeny computations if needed.

Since these modifications generally produce more costly group action computations,
the authors also provide several optimizations that limit the
slow-down in their algorithm to a factor of $3.10$ compared to \cite{MR18}. 
These include
the Elligator 2 map of~\cite{BHKL13} and \cite{BLMP18},
multiple batches for isogeny computation (SIMBA),
and sample the exponents $e_i$ from  intervals of different sizes
depending on $\ell_i$.

\subsection{The Onuki--Aikawa--Yamazaki--Takagi constant-time algorithm}

Still assuming that the attacker can perform only power consumption
analysis and timing attacks, Onuki, Aikawa, Yamazaki and Takagi proposed
a faster constant-time version of CSIDH in~\cite{OAYT19}.

The key idea is to use two points to evaluate the action of an ideal,
one in $\ker(\pi-1)$ (i.e., in $E(\Fp)$) and one in $\ker(\pi+1)$
(i.e., in $E(\Fq)$ with $x$-coordinate in $\Fp$).
This allows them to avoid timing attacks,
while keeping the same primes and exponent range $[-5, 5]$
as in the original CSIDH algorithm. 
Their algorithm also employs dummy isogenies
to mitigate some power analysis attacks, as in~\cite{MCR18}.
With these improvements, they achieve a speed-up of $27.35 \%$
compared to \cite{MCR18}.

We include pseudo-code for the algorithm of~\cite{OAYT19}
in Algorithm~\ref{alg:OAYT},
to serve both as a reference for a discussion of a subtle leak
in~\S\ref{sec:repair}
and also as a departure point for our dummy-free algorithm
in~\S\ref{sec:fault}.

\begin{algorithm}[ht]
	\caption{The Onuki--Aikawa--Yamazaki--Takagi CSIDH algorithm
        for supersingular curves over $\Fp$,
        where $p = 4 \prod_{i=1}^n \ell_i -1$.
        The ideals 
        $\frakl_i = (\ell_i, \pi - 1)$, where \(\pi\) 
        maps to the $p$-th power Frobenius endomorphism on each curve,
        and the exponent bound vector~\((m_1,\ldots,m_n)\),
        are system parameters.
        This algorithm computes exactly \(m_i\) isogenies for each~\(\ell_i\).
    }
    \label{alg:OAYT}
    \KwIn{A supersingular curve \(E_A \colon y^2 = x^3 + Ax^2 + x\) over
    \(\Fp\),
    and an integer exponent vector $(e_1, \dots, e_n)$ 
    with each \(e_i \in [-m_i,m_i]\).}
    \KwOut{$E_B \colon y^2 = x^3 + Bx^2 + x$ 
    such that $E_B = \frakl_1^{e_1} \ast \cdots \ast \frakl_n^{e_n} \ast E_A$.}
	$(e_1', \ldots, e_n') \gets (m_i - |e_1|, \ldots, m_i - |e_n|)$
	\tcp*{Number of dummy computations}
	\(E_B \gets E_A\)
	\;
    \While{some $e_i \ne 0$ or $ e_i' \ne 0$}{
		$S \gets \{i \mid e_i \ne 0$ or $e_i' \ne 0\}$
		\;
		$k \gets \prod_{i \in S} \ell_i$
        \;
    	$u \gets \texttt{Random}\big( \big\{2, \ldots, \frac{p-1}{2}\big\}\big)$
        \label{alg:OAYT:random-call}
   	    \;
    	\( (T_-, T_+) \gets \texttt{Elligator}(E_B,u)\)
        \label{alg:mcr-elligator} 
    	\tcp*{\(T_- \in E_B[\pi-1]\) and \(T_+ \in E_B[\pi+1]\) }
        $(P_0, P_1) \gets \big([(p+1)/k]T_+, [(p+1)/k]T_-\big)$
        \;
        \For{$i \in S$}{
            $s \gets \text{sign}(e_i)$
            \tcp*{Ideal \(\frakl_i^s\) to be used}
			$Q \gets [k/\ell_i]P_{\frac{1 - s}{2}}$
            \label{alg:OAYT:compute-Q}
            \tcp*{Secret kernel point generator}
			$P_{\frac{1+s}{2}} \gets [\ell_i]P_{\frac{1+s}{2}}$
			\tcp*{Secret point to be multiplied}
            \If{$Q \ne \infty$}{
                \label{alg:OAYT:bad-if}
                \uIf{$e_i\ne 0$}{
                  \label{alg:OAYT:dummy-if}
                    $(E_B,\varphi) \gets \texttt{QuotientIsogeny}(E_B,Q)$
                    \;
                    $(P_0,P_1) \gets \big(\varphi(P_0),\varphi(P_1)\big)$
                    \;
                    $e_i \gets e_i - s$.
                }
                \Else{ 
					$E_B \gets E_B$; $P_{\frac{1-s}{2}} \gets [\ell_i]P_\frac{1-s}{2}$; $e_i' \gets e_i' - 1$
                    \label{alg:OAYT:dummy}
                    \tcp*{Dummies}
                }
            }
			$k \gets k/\ell_i$
        }
    }
    \Return{$B$}
\end{algorithm}

\section{
    Repairing constant-time versions based on Elligator
}
\label{sec:repair}


Both~\cite{MCR18} and~\cite{OAYT19} use the Elligator 2 map to sample
a random point on the current curve $E_A$ in
step~\ref{alg:mcr-elligator} of Algorithm~\ref{alg:OAYT}. %
Elligator takes as input a random field element
$u\in\{2,\dots,\frac{p-1}{2}\}$ and the Montgomery $A$-coefficient
from the current curve and returns a pair of points in $E_A[\pi - 1]$
and $E_A[\pi + 1]$ respectively.

To avoid a costly inversion of $u^2 - 1$, instead of sampling
$u$ randomly, Meyer, Campos and Reith%
\footnote{Presumably, Onuki \emph{et al.} do the same, however their
  exposition is not clear on this point, and we do not have access to
  their code.} %
follow~\cite{BLMP18} and precompute a set of ten pairs
$(u,(u^2-1)^{-1})$; they try them in order until one that produces a
point $Q$ passing the test in Step~\ref{alg:OAYT:bad-if} is found.
When this happens, the algorithm moves to the next curve, and
Elligator can keep on using the next precomputed value of $u$,
going back to the first value when the tenth has been reached. 
This is a major departure from~\cite{BLMP18}, where
\emph{all} precomputed values of $u$ are tried 
\emph{for each isogeny computation}, 
and the algorithm
succeeds if at least one passes the test. %
And indeed the implementation of~\cite{MCR18} leaks information on the
secret via the timing channel:%
\footnote{The Elligator optimization is described in \S5.3
  of~\cite{MCR18}. The unoptimized constant-time version described in
  Algorithm~2 therein is not affected by this problem.} %
since Elligator uses no randomness for $u$, its output only depends on
the $A$-coefficient of the current curve, which itself depends on the
secret key; but the running time of the algorithm varies and, not
being correlated to $u$, it is necessarily correlated to $A$ and thus
to the secret.

Fortunately this can be easily fixed by (re)introducing randomness in
the input to Elligator.  %
To avoid field inversions, we use a projective variant: given
$u\ne0,1$ and assuming $A\ne0$, we write $V = (A : u^2 -1)$, and we
want to determine whether $V$ is the abscissa of a projective point on
$E_A$. %
Plugging $V$ into the homogeneous equation
\[E_A : Y^2Z^2 = X^3Z + AX^2Z^2 + XZ^3\]
gives
\[Y^2(u^2-1)^2 = \bigl((A^2u^2 + (u^2-1)^2\bigr)A(u^2-1).\] %
We can test the existence of a solution for $Y$ by computing the
Legendre symbol of the right hand side: if it is a square, the points
with projective $XZ$-coordinates
\begin{align*}
  T_+ &= (A:u^2-1),
    &
  T_- &= (-Au^2:u^2-1)
\end{align*}
are in $E_A[\pi-1]$ and $E_A[\pi+1]$ respectively, otherwise their
roles are swapped. %

We are left with the case $A=0$. %
Following~\cite{BLMP18}, Meyer, Campos and Reith precompute once and
for all a pair of generators $T_+,T_-$ of $E_0[\pi-1]$ and
$E_0[\pi+1]$, and output those instead of random points. %
This choice suffers from a similar issue to the previous one: because
the points are output in a deterministic way, the running time of the
whole algorithm will be correlated to the number of times the curve
$E_0$ is encountered during the isogeny walk.

In practice, $E_0$ is unlikely to ever be encountered in a random
isogeny walk, except as the starting curve in the first phase of a key
exchange, thus this flaw seems hard to exploit. %
Nevertheless, we find it not significantly more expensive to use a
different approach, also suggested in~\cite{BLMP18}: with $u\ne0$,
only on $E_0$, we define the output of Elligator as
$T_+=(u:1),T_-=(-u:1)$ when $u^3+u$ is a square, and we swap the
points when $u^3+u$ is not a square.

We summarize our implementation of Elligator in
Algorithm~\ref{alg:elligator}, generalizing it to the case of
Montgomery curves represented by projective coefficients (see also
Section~\ref{sec:montg}).

\begin{algorithm}
    \caption{Constant-time projective Elligator}
    \label{alg:elligator}
    \KwIn{A supersingular curve $E_{(A':C')}: C'y^2=C'x^3+A'x^2+C'x$ over
    \(\Fp\), and an element $u\in\{2,\dots,\frac{p-1}{2}\}$.}%
    \KwOut{A pair of points $T_+\in E_{(A':C')}[\pi-1]$ and $T_-\in E_{(A':C')}[\pi+1]$.}
    $t \gets A'\big((u^2-1)u^2{A'}^2C' + ((u^2-1)C')^3\big)$
    \;
    $a \gets \texttt{isequal}(t,0)$
    \tcp*{$t=0$ iff $A'=0$}
    $\alpha, \beta \gets 0, u$
    \;
    $\texttt{cswap}(\alpha, \beta,a)$
    \tcp*{$\alpha=0$ iff $A'\ne0$}
    $t' \gets t + \alpha(u^2+1)$
    \tcp*{$t'\ne0$}
    $T_+ \gets (A' + \alpha C'(u^2-1) : C'(u^2 - 1))$
    \;
    $T_- \gets (-A'u^2 - \alpha C'(u^2-1) : C'(u^2 - 1))$
    \;
    $b \gets \texttt{Legendre\_symbol}(t',p)$
    \tcp*{$b=\pm1$}
    $c \gets \texttt{isequal}(b,-1)$
    \;
    $\texttt{cswap}(T_+,T_-,c)$\;
    \Return $(T_+,T_-)$
    \;
\end{algorithm}


We can now prove that Algorithm~\ref{alg:OAYT} runs in time
independent from the secret vector $(e_1,\dots,e_n)$. %
First, assume that Elligator always returns points of order $p+1$,
then it is clear that the condition in Line~\ref{alg:OAYT:bad-if} is
always true, and thus that the outer loop runs exactly $\max(m_i)$
times. %
The only branching depending on secrets is then the one at
Line~\ref{alg:OAYT:dummy-if}, however the two branches take exactly
the same time, thanks to the dummy computations.

In general, we cannot assume that Elligator always returns full order
points, however, under reasonable heuristics experimentally verified
in~\cite{BLMP18},
\[\Prob\{[(p+1)/\ell_i]T_+ = \infty\} = \Prob\{[(p+1)/\ell_i]T_- =
  \infty\} = 1/\ell_i\] %
for any prime $\ell_i$ and any curve $E_B$.%
\footnote{Note that the joint probability of $T_+$ and $T_-$ having
  order divisible by $\ell_i$ is not independent of $E_B$, however
  this will not be a problem in our algorithms.} %
Then, even though the value of $Q$ in Line~\ref{alg:OAYT:compute-Q}
depends on the sign $s$ of the secret exponent~\(e_i\), the
probability that the test in Line~\ref{alg:OAYT:bad-if} passes is
independent of all secrets. %

\section{
    Optimizing constant-time implementations
}
\label{sec:optimize}

In this section we propose several optimizations that are compatible
with both non-constant-time and constant-time implementations of
CSIDH.

\subsection{Isogeny and point arithmetic on twisted Edwards curves}\label{sec:edwards}

In this subsection, we present efficient formulas 
in twisted-Edwards coordinates 
for four fundamental operations:
point addition, 
point doubling, 
isogeny computation (as presented in~\cite{Moody11}; cf.~\S\ref{sec:csidh}), 
and 
isogeny evaluation 
(\emph{i.e.} computing the image of a point under an isogeny).
Our approach obtains a modest but still noticeable improvement with 
respect to previous proposals based on Montgomery representation,
or hybrid strategies that propound combinations of Montgomery and twisted-Edwards 
representations~\cite{CGF08,KYKHP18,KYKPH18,KY0H19,MRC17}.

Castryck, Galbraith, and Farashahi~\cite{CGF08} proposed using a hybrid representation
to reduce the cost of point doubling on certain Montgomery curves,
by exploiting the fact that converting between Montgomery and twisted
Edwards models can be done at almost no cost. 
In~\cite{MRC17}, Meyer,
Reith and Campos considered using twisted Edwards formulas for computing
isogeny and elliptic curve arithmetic, but concluded that a pure
twisted-Edwards-only approach would not be advantageous in the context of SIDH.
Bernstein, Lange, Martindale, and Panny observed in~\cite{BLMP18}
that the conversion from Montgomery XZ coordinates to twisted
Edwards YZ coordinates occurs naturally during the Montgomery ladder.
Kim, Yoon, Kwon, Park, and Hong presented a hybrid model in~\cite{KYKPH18} 
using Edwards and Montgomery modelds for isogeny 
computations and point arithmetic, respectively; 
in~\cite{KYKHP18} and~\cite{KY0H19}, they suggested computing isogenies using a modified twisted Edwards representation that introduces a fourth coordinate $w$. 
   
To the best of our knowledge, the quest for more efficient
elliptic curve and isogeny arithmetic than that offered by pure Montgomery
and twisted-Edwards-Montgomery representations remains an open problem. As a step forward in this direction, Moody and Shumow~\cite{Moody11} showed that 
when dealing with isogenies of odd degree $d = 2\ell - 1$ with $\ell \ge 2,$ twisted Edwards representation offers a cheaper formulation for isogeny computation 
than the corresponding one using Montgomery curves; nevertheless, they
did not address the problem of getting a cheaper twisted Edwards formulation for 
the isogeny evaluation operation.

\subsubsection{Montgomery curves}\label{sec:montg}
A Montgomery curve~\cite{Montgomery87} is defined by the equation 
$E_{A,B}: By^2 = x^3 + Ax^2 + x$, such that $B \ne 0$ and $A^2 \ne 4$
(we often write \(E_{A}\) for \(E_{A,1}\)).
We refer to \cite{CS17} for a survey on Montgomery curves. 
When performing isogeny computations and evaluations, 
it is often more convenient to represent the constant
$A$ in the projective space $\mathbb{P}^1$ as $(A': C'),$ such that $A = A'/C'.$ 
Montgomery curves are attractive because they are exceptionally well-suited
to performing the differential point addition operation
which computes \(x(P+Q)\) from \(x(P)\), \(x(Q)\), and \(x(P-Q)\).
Equations~\eqref{eq:xzdbl} and \eqref{eq:xzadd}
describe the
differential point doubling and  addition operations proposed by
Montgomery in~\cite{Montgomery87}:
\begin{align}
    X_{[2]P} &= C_{24}(X_{P}+ Z_{P})^2 (X_{P} - Z_{P})^2,\label{eq:xzdbl}\\
    Z_{[2]P} &= ((X_{P}+ Z_{P})^2  - (X_{P} - Z_{P})^2)\cdot\nonumber \\
             &\qquad (C_{24}(X_{P} - Z_{P})^2  + A_{24p}((X_{P}+ Z_{P})^2  - (X_{P} - Z_{P})^2))\nonumber 
    \intertext{where $A_{24p} = A + 2C$ and $C_{24} = 4C$, and}
%
    X_{P+Q} &= Z_{P-Q}\left[(X_P - Z_P)(X_Q + Z_Q) + (Z_P + Z_P)(X_Q - Z_Q) \right]^2 \label{eq:xzadd}\\
    Z_{P+Q} &= X_{P-Q}\left[(X_P - Z_P)(X_Q + Z_Q) - (Z_P + Z_P)(X_Q - Z_Q) \right]^2 \nonumber 
\end{align}
    
Montgomery curves can be used to efficiently compute isogenies
using V\'elu's formulas~\cite{Vel71}. 
Suppose we want the image of a point \(Q\)
under an \(\ell\)-isogeny~\(\phi\), where \(\ell = 2k+1\).
For each \(1 \le i \le k\) we let $(X_i: Z_i) = x([i]P)$,
where \(\langle{P}\rangle = \ker\phi\). 
Equation~\eqref{eq:xzeval} 
computes \((X':Z') = x(\phi(Q))\) from \((X_Q:Z_Q) = x(Q)\).
\begin{align}
        X' &= X_P\Big(\prod_{i=1}^{k}\big[(X_Q - Z_Q)(X_i + Z_i) +
        (Z_Q + Z_Q)(X_i - Z_i) \big] \Big)^2 \label{eq:xzeval}\\
        Z' &= Z_P\Big(\prod_{i=1}^{k}\big[(X_Q - Z_Q)(X_i + Z_i) -
        (Z_Q + Z_Q)(X_i - Z_i) \big] \Big)^2 \nonumber 
\end{align}

\subsubsection{Twisted Edwards curves}
In~\cite{BernsteinBJLP08}
we see that every Montgomery curve \(E_{A,B}: By^2 = x^3 + Ax^2 + x\)
is birationally equivalent to a twisted Edwards curve
\(E_{a,d} : ax^2 + y^2 = 1 + dx^2y^2\);
the curve constants are related by
\[
    (A,B) = \left(\frac{2(a+d)}{a-d},\frac{4}{a-d}\right)
    \quad
    \text{and}
    \quad
    (a,d) = \left(\frac{A+2}{B}, \frac{A-2}{B}\right)
    \,,
\]
and the rational maps
\(\phi: E_{a,d} \to E_{A,B}\)
and
\(\psi: E_{A,B} \to E_{a,d}\)
are defined by
\begin{align}
    \phi: (x,y) & \longmapsto \left((1+y)/(1-y), (1+y)/(1-yx)\right)
    \,,
    \nonumber 
    \\
    \psi: (x,y) & \longmapsto \left(x/y, (x-1)/(x+1)\right)
    \,.
    \label{eq:monttoed}
\end{align}
Rewriting this relationship for Montgomery curves
with projective constants,
\(E_{a,d}\)
is equivalent to the Montgomery curve
$E_{(A: C)} = E_{A/C,1}$ 
with constants
\begin{align*}
    A_{24p} & := A + 2C = a \,,
    &
    A_{24m} & := A - 2C = d \,,
    &
    C_{24} & := 4C = a - d \,.
\end{align*}
To avoid notational ambiguities, 
we write $(Y_P: T_P)$ for the $\mathbb{P}^1$ projection 
of the $y$-coordinate of the point $P \in E_{a,d}$.
Let $P\in E_{(A:C)}$. 
In projective coordinates, the map \(\psi\) of~\eqref{eq:monttoed} becomes
\begin{align}
    \psi: (X_P:Z_P) \longmapsto (Y_P:T_P) = (X_P - Z_P : X_P + Z_P) \label{eq:xzmonttoyted}
\end{align}
Comparing~\eqref{eq:xzmonttoyted} with~\eqref{eq:xzdbl} 
reveals that $Y_P$ and $T_P$ appear in the doubling formula,
so we can substitute them at no cost.
Replacing
$A_{24p}$ and $C_{24}$ 
with their twisted Edwards equivalents $a$ and $e=a-d$, respectively,
we obtain a doubling formula for twisted Edwards $YT$ coordinates:
\begin{align*}
    Y_{[2]P} &= e\cdot Y_P^2\cdot T_P^2 - (T_P^2  - Y_P^2)\cdot (eY_P^2 + a(T_P^2  - Y_P^2)),
    \\
    T_{[2]P} &= e\cdot Y_P^2\cdot T_P^2 + (T_P^2  - Y_P^2)\cdot (eY_P^2 + a(T_P^2  - Y_P^2)).
\end{align*}

Similarly, the coordinates $Y_P, T_P, Y_Q, T_Q, Y_{P-Q}$ and $T_{P-Q}$
appear in~\eqref{eq:xzadd},
and thus we
derive differential addition formulas for twisted Edwards
coordinates:
\begin{align*}
    Y_{P+Q} &= (T_{P-Q} - Y_{P-Q})(Y_P T_Q + Y_Q Z_P )^2 - (T_{P-Q} + Y_{P-Q})(Y_P T_Q - Y_Q Z_P)^2,
    \\
    T_{P+Q} &= (T_{P-Q} - Y_{P-Q})(Y_P T_Q + Y_Q Z_P )^2 + (T_{P-Q} + Y_{P-Q})(Y_P T_Q - Y_Q Z_P)^2.
\end{align*}
The computational costs of doubling and differential addition
are $4\textbf{M} + 2\textbf{S} + 4\textbf{A}$ (the same as evaluating~\eqref{eq:xzdbl})
and $4\textbf{M} + 2\textbf{S} + 6\textbf{A}$ (the same as~\eqref{eq:xzadd}),
respectively.
    
The Moody--Shumow formulas for isogeny computation~\cite{Moody11}
are given in terms of twisted Edwards $YT$-coordinates. 
It remains to derive a twisted Edwards $YT$-coordinate
isogeny-evaluation formula
for \(\ell\)-isogenies where \(\ell = 2k+1\).
We do this by applying the map in~\eqref{eq:xzmonttoyted} 
to~\eqref{eq:xzeval}, which yields
\begin{align*}
    Y'
    & =
    (T_{P-Q} + Y_{P-Q}) \cdot \Big(\prod_{i=1}^k \left[ T_{Q}Y_{[i]P} +
    Y_{Q}T_{[i]P}\right]\Big)^2
    \\
    &
    \qquad - (T_{P-Q} - Y_{P-Q}) \cdot \Big(\prod_{i=1}^k \left[
        T_{Q}Y_{[i]P} - Y_{Q}T_{[i]P}\right]\Big)^2
    \,,
    \\
    T'
    & =
    (T_{P-Q} + Y_{P-Q}) \cdot \Big(\prod_{i=1}^k \left[ T_{Q}Y_{[i]P} +
    Y_{Q}T_{[i]P}\right]\Big)^2
    \\
    &
    \qquad + (T_{P-Q} - Y_{P-Q}) \cdot \Big(\prod_{i=1}^k \left[
        T_{Q}Y_{[i]P} - Y_{Q}T_{[i]P}\right]\Big)^2
    \,.
\end{align*}
 
The main advantage of the approach outlined here is that by only using points given in $YT$ coordinates, we can compute point doubling, point addition and
isogeny construction and evaluation at a lower computational cost.
Indeed, isogeny evaluation in $XZ$ costs $4k\textbf{M} + 2\textbf{S} +
6k\textbf{A}$, whereas the above 
$YT$ coordinate formula costs $4k\textbf{M} + 2\textbf{S} + (2k + 4)\textbf{A}$,
thus saving $4k - 4$ field additions.

\subsection{Addition chains for a faster scalar multiplication} 

Since the coefficients in CSIDH scalar multiplications are always known in advance
(they are essentially system parameters), 
there is no need to hide them by using constant-time scalar multiplication
algorithms such as the classical Montgomery ladder.
Instead, we can use shorter differential addition chains.\footnote{%
    A differential addition chain is an addition chain such that for
    every chain element $c$ computed as $a + b$, the difference $a-b$ is
    already present in the chain.
}

In the CSIDH group action computation, any given scalar $k$ is the
product of a subset of the collection of the $74$ small primes
\(\ell_i\) dividing $\frac{p+1}{4}$. 
We can take advantage of this structure to use shorter differential
addition chains than those we might derive for general
scalars of a comparable size.
First, we pre-computed the shortest differential
addition chains for each one of the small primes $\ell_i$.
One then computes the scalar multiplication operation $[k]P$ 
as the composition of the differential addition chains for each prime
$\ell$ dividing $k$. 

Power analysis on the coefficient computation might reveal the degree of
the isogeny that is currently being computed, but, since we compute
exactly one $\ell_i$-isogeny for each $\ell_i$ per loop, this does not
leak any secret information.

This simple trick allows us to compute scalar multiplications $[k]P$
using differential addition chains of length roughly $1.5\lceil\log_2(k)\rceil$.
This yields a saving of about 25\% compared with the cost of the classical Montgomery ladder.

\section{
    Removing dummy operations for fault-attack resistance
}
\label{sec:fault}

The use of dummy operations in the previous constant-time algorithms
implies that the attacker can obtain information on the secret key
by injecting faults into variables during the computation.
If the final result is correct, then she knows that the fault was injected in
a dummy operation; if it is incorrect, then the operation was real.
For example, if one of the values in Line~\ref{alg:OAYT:dummy} of Algorithm~\ref{alg:OAYT}
is modified without affecting the final result, then the adversary
learns whether the corresponding exponent \(e_i\) was zero at that point.

Fault injection attacks have been considered 
in the context of SIDH (\cite{GW17}, \cite{Ti17}), 
but to the best of our knowledge,
they have not been studied yet on dummy operations in the context of CSIDH.
Below we propose an approach to constant-time CSIDH
without dummy computations,
making every computation essential for a correct final result.
This gives us some natural resistance to faults,
at the cost of approximately a twofold slowdown.

Our approach to avoiding fault-injection attacks is to change the
format of secret exponent vectors $(e_1,\dots,e_n)$. %
In both the original CSIDH and the Onuki \emph{et al.} variants, the
exponents $e_i$ are sampled from an integer interval $[-m_i,m_i]$ centered
in $0$.
For naive CSIDH, evaluating the action of \(\frakl_i^{e_i}\) 
requires evaluating between \(0\) and \(m_i\) isogenies,
corresponding to either the ideal \(\frakl_i\) (for positive \(e_i\))
or \(\frakl_i^{-1}\) (for negative \(e_i\)).
If we follow the approach of~\cite{OAYT19},
then we must also compute \(m_i - |e_i|\) dummy \(\ell_i\)-isogenies
to ensure a constant-time behaviour.

For our new algorithm, the exponents $e_i$ are uniformly sampled from
sets
\begin{equation*}
  \Set(m_i) = \{e \;|\; e = m_i \bmod 2 \text{ and } \lvert e\rvert \le m_i \},
\end{equation*}
i.e., centered intervals containing only even or only odd integers. %
The interesting property of these sets is that a vector drawn from
$\Set(m)^n$ can always be rewritten (in a non-unique way) as a sum of
$m$ vectors with entries $\{-1,+1\}$ (i.e., vectors in $\Set(1)^n$). %
But the action of a vector drawn from $\Set(1)^n$ can clearly be
implemented in constant-time without dummy operations: for each
coefficient $e_i$, we compute and evaluate the isogeny associated to
$\frakl_i$ if $e_i=1$, or the one associated to $\frakl_i^{-1}$ if
$e_i=-1$. %
Thus, we can compute the action of vectors drawn from $\Set(m)^n$ by
repeating $m$ times this step.

More generally, we want to evaluate the action of vectors
$(e_1,\ldots,e_n)$ drawn from
$\Set(m_1)\times\cdots\times\Set(m_n)$. %
Algorithm~\ref{alg:idealized} achieves this in constant-time and
without using dummy operations. %
The outer loop at line~\ref{alg:idealized:outer} is repeated exactly
$\max(m_i)$ times, but the inner ``if'' block at
line~\ref{alg:idealized:if} is only executed $m_i$ times for each $i$;
it is clear that this flow does not depend on secrets. %
Inside the ``if'' block, the coefficients $e_i$ are implicitly
interpreted as
\begin{equation*}
    \lvert e_i\rvert = \underbrace{1 + 1 + \cdots + 1}_{e_i\text{ times}} +
    \underbrace{(1 - 1) - (1 - 1) + (1 - 1) - \cdots}_{m_i - e_i\text{ times}},
\end{equation*}
i.e., the algorithm starts by acting by
$\frakl_i^{\texttt{sign}(e_i)}$ for $e_i$ iterations, then alternates
between $\frakl_i$ and $\frakl_i^{-1}$ for $m_i-e_i$ iterations. %
We assume that the $\texttt{sign}:\mathbb{Z}\to\{\pm1\}$ operation is
implemented in constant time, and that $\texttt{sign}(0)=1$. %
If one is careful to implement the isogeny evaluations in
constant-time, then it is clear that the full algorithm is also
constant-time.

\begin{algorithm}[h]
  \caption{An idealized dummy-free constant-time evaluation of the CSIDH
    group action.}
  \label{alg:idealized}
  \KwIn{Secret vector $(e_1,\dots,e_n)\in\Set(m_1)\times\cdots\times\Set(m_n)$}
  $(t_1, \ldots, t_n) \gets (\texttt{sign}(e_1), \ldots, \texttt{sign}(e_n))$
  \tcp*{Secret}
  $(z_1, \ldots, z_n) \gets (m_1, \ldots, m_n)$
  \tcp*{Not secret}
  \While{\label{alg:idealized:outer} some $z_i\ne 0$}{
    \For{$i\in\{1,\dots,n\}$}{
      \If{\label{alg:idealized:if} $z_i > 0$}{
        Act by $\frakl_i^{t_i}$
        \;
        $b = \texttt{isequal}(e_i,0)$
        \;
        $e_i \gets e_i - t_i$
        \;
        $t_i \gets (-1)^b \cdot t_i$
        \tcp*{Swap sign when $e_i$ has gone past $0$}
        $z_i \gets z_i - 1$
        \;
      }
    }
  }
\end{algorithm}

However, Algorithm~\ref{alg:idealized} is only an idealized version of
the CSIDH group action algorithm. %
Indeed, like in~\cite{MCR18,OAYT19}, it may happen in some iterations
that Elligator outputs points of order not divisible by $\ell_i$, and
thus the action of $\frakl_i$ or $\frakl_i^{-1}$ cannot be computed in
that iteration. %
In this case, we simply skip the loop and retry later: this translates
into the variable $z_i$ not being decremented, so the total number of
iterations may end up being larger than $\max(m_i)$. %
Like in Section~\ref{sec:repair}, if the input value $u$ fed to
Elligator is random, its output is uncorrelated to secret values%
\footnote{Assuming the usual heuristic assumptions on the distribution
  of the output of Elligator, see~\cite{MCR18}.}, %
and thus the fact that an iteration is skipped does not leak
information on the secret. %
The resulting algorithm is summarized in
Algorithm~\ref{alg:DummyFree}.

To maintain the security of standard CSIDH, the bounds $m_i$ must
be chosen so that the key space is at least as large. %
For example, the original implementation~\cite{CLMPR18} samples
secrets in $[-5,5]^{74}$, which gives a key space of size $11^{74}$;
hence, to get the same security we would need to sample secrets in
$\Set(10)^{74}$. %
But a constant-time version of CSIDH \emph{à la} Onuki \emph{et al.}
only needs to evaluate five isogeny steps per prime $\ell_i$, whereas
the present variant would need to evaluate ten isogeny steps. %
We thus expect an approximately twofold slowdown for this variant
compared to Onuki \emph{et al.}, which is confirmed by our experiments.

\begin{algorithm}
    \caption{
        Dummy-free randomized constant-time CSIDH class group action 
        for supersingular curves over $\Fp$, where $p = 4 \prod_{i=1}^n \ell_i -1$.
        The ideals 
        $\frakl_i = (\ell_i, \pi - 1)$, where \(\pi\) 
        maps to the $p$-th power Frobenius endomorphism on each curve,
        and the vector \((m_1,\ldots,m_n)\) of exponent bounds,
        are system parameters.
        This algorithm computes exactly \(m_i\) isogenies for each ideal \(\frakl_i\).
    }
	\label{alg:DummyFree}
    \KwIn{A supersingular curve $E_{A}$ over \(\Fp\),
        and an exponent vector $(e_1, \dots, e_n)$ with each \(e_i \in [-m_i, m_i]\) and 
        \(e_i \equiv m_i \pmod 2\).}
    \KwOut{$E_B = \frakl_1^{e_1} \ast \cdots \ast \frakl_n^{e_n} \ast E_A$.}
    \((t_1, \ldots, t_n) \gets \left(\frac{\texttt{sign}(e_1) + 1}{2}, \ldots, \frac{\texttt{sign}(e_n) + 1}{2}\right)\)
    \tcp*{Secret}
    \((z_1, \ldots, z_n) \gets (m_1, \ldots, m_n)\)
    \tcp*{Not secret}
    \(E_B \gets E_A\)\;
    \While{some \(z_i \neq 0\)}{
    	$u \gets \texttt{Random}\big( \big\{2, \ldots, \frac{p-1}{2}\big\}\big)$
        \label{alg:DummyFree:random-call}
   	    \;
    	\( (T_1, T_0) \gets \texttt{Elligator}(E_B,u)\)
    	\tcp*{\(T_1 \in E_B[\pi-1]\) and \(T_0 \in E_B[\pi+1]\) }
        $(T_0,T_1) \gets ([4]T_0,[4]T_1)$
        \tcp*{Now $T_0, T_1 \in E_B\left[\prod_i\ell_i\right]$}
        \For{$i \in \{1,\ldots,n\}$}{
        	\If{\(z_i \neq 0\)}{
                \label{alg:DummyFree:branch-1}
	            $(G_0,G_1) \gets (T_0, T_1)$
    	        \;
                \(\texttt{cswap}(G_0, G_1, t_i)\)
                \tcp*{Secret kernel point generator: \(G_0\)}
                \(\texttt{cswap}(T_0, T_1, t_i)\)
                \tcp*{Secret point to be multiplied: \(T_1\)}
            	\For{$j \in \{i+1,\ldots,n\}$}{
            		$G_0 \gets [\ell_j]G_0$
	            }
	            \If{\label{alg:oayt-leaky} \(G_0 \neq \infty\)}{
                    \label{alg:DummyFree:branch-2}
    	    	    $(E_B, \phi) \gets \texttt{QuotientIsogeny}(E_B, G_{0})$
    	    	    \;
	        	    $(T_0,T_1) \gets \big(\phi(T_0),\phi(T_1)\big)$
    	        	\;
    	    	    \(b \gets \texttt{isequal}(e_i, 0)\)
    	    	    \;
		            \(e_i \gets e_i + (-1)^{t_i} \)
    		        \;
        		    \(t_i \gets t_i \oplus b\)
        		    \;
        		    \(z_i \gets z_i - 1\)
            	}
                \(T_{1} \gets [\ell_i]T_{1}\)
                \;
                \(\texttt{cswap}(T_0, T_1, t_i)\)
                \;
            }
        }
    }
    \Return{$B$}
\end{algorithm}

\section{
    Derandomized CSIDH algorithms
}
\label{sec:derandomized}

As we stressed in Section~\ref{sec:repair}, all of the algorithms
presented here depend on the availability of high-quality randomness
for their security. Indeed, the input to Elligator must be randomly
chosen to ensure that the total running time is uncorrelated to the
secret key. %
Typically, this would imply the use of a PRNG seeded with high quality
true randomness that must be kept secret. %
An attack scenario where the attacker may know the output of the PRNG,
or where the quality of PRNG output is less than ideal, therefore
degrades the security of all algorithms.
This is true even when the secret was generated with a high-quality PRNG
if the keypair is static, and the secret key is then used by an
algorithm with low-quality randomness.

We can avoid this issue completely if points of order
$\prod \ell_i^{|m_i|}$, where $|m_i|$ is the maximum possible exponent
(in absolute value) for $\ell_i$, are available from the start. %
Unfortunately this is not possible with standard CSIDH, because such
points are defined over field extensions of exponential degree. 

Instead, we suggest modifying CSIDH as follows.  First, we take a
prime $p = 4 \prod_{i=1}^{n} \ell_i - 1$ such that
$\lceil n\log_2(3) \rceil = 2\lambda$, where $\lambda$ is a security
parameter, and we restrict to exponents of the private key sampled
from $\{-1, 0, 1 \}$.  Then, we compute two points of order $(p+1)/4$
on the starting public curve, one in $\ker(\pi -1)$ and the other in
$\ker(\pi + 1)$, where $\pi$ is the Frobenius endomorphism. %
This computation involves no secret information and can be implemented
in variable-time; furthermore, if the starting curve is the initial
curve with $A=0$, or a public curve corresponding to a long term
secret key, these points can be precomputed offline and attached to
the system parameters or the public key. %
We also remark that even for ephemeral public keys, a point of order
$p+1$ must be computed anyway for key validation purposes, and thus
this computation only slows down key validation by a factor of two.

Since we have restricted exponents to $\{-1, 0, 1\}$, every
$\ell_i$-isogeny in Algorithm~\ref{alg:OAYT} can be computed using
only (the images of) the two precomputed points. There is no possibility
of failure in the test of Line~\ref{alg:OAYT:bad-if}, and no need
to sample any other point.

We note that this algorithm still uses dummy operations. %
If fault-injection attacks are a concern, the exponents can be further
restricted to $\{-1,1\}$, and the group action evaluated as in (a
stripped down form of) Algorithm~\ref{alg:DummyFree}. %
However this further increases the size of $p$, as $n$ must now be equal
to $2\lambda$.

This protection comes at a steep price: at the 128 bits security
level, the prime $p$ goes from 511 bits to almost 1500. %
The resulting field arithmetic would be considerably slower, although
the global running time would be slightly offset by the smaller number
of isogenies to evaluate.

On the positive side, the resulting system would have 
much stronger quantum security. %
Indeed, the best known quantum attacks are exponential in the size of
the key space ($\approx 2^{2\lambda}$ here), but only subexponential
in~$p$ (see~\cite{CJS10,DFKS18,CLMPR18}). Since our modification more
than doubles the size of $p$ without changing the size of the key space,
quantum security is automatically increased. %
For this same reason, for security levels beyond NIST-1 (64 quantum
bits of security), the size of $p$ increases more than linearly in
$\lambda$, and the variant proposed here becomes natural. %
Finally, parameter sets with a similar imbalance between the size of
$p$ and the security parameter $\lambda$ have already been considered
in the context of isogeny based signatures~\cite{SeaSign}, where they
provide tight security proofs in the QROM.

Hence, while at the moment this costly modification of CSIDH may seem
overkill, we believe further research is necessary to try and bridge
the efficiency gap between it and the other side-channel protected
implementations of CSIDH.

\section{
    Experimental results
}

Tables~\ref{tab:ops-revised}
and~\ref{tab:cc-revised}
summarize our experimental results,
and compare our algorithms
with those of~\cite{CLMPR18}, \cite{MCR18}, and~\cite{OAYT19}.
Table~\ref{tab:ops-revised}
compares algorithms in terms of elementary field operations,
while Table~\ref{tab:cc-revised}
compares cycle counts of C implementations.
All of our experiments were ran on a Intel(R) Core(TM) i7-6700K CPU 4.00GHz
machine with 16GB of RAM.  Turbo boost was disabled.
The software environment was 
the Ubuntu 16.04 operating system and \texttt{gcc} version 5.5.

In all of the algorithms considered here
(except the original~\cite{CLMPR18}),
the group action is evaluated using the SIMBA method 
(Splitting Isogeny computations into Multiple BAtches) 
proposed by Meyer, Campos, and Reith in~\cite{MCR18}. 
Roughly speaking, SIMBA-\(m\)-\(k\) partitions the set of primes \(\ell_i\)
into \(m\) disjoint subsets \(S_i\) (batches) of approximately the same size.
SIMBA-\(m\)-\(k\) proceeds by computing isogenies for each batch
\(S_i\); after \(k\) steps, the unreached primes \(\ell_i\) 
from each batch are merged.

\paragraph{Castryck et al.}
We used the reference CSIDH implementation made available
for download by the authors of~\cite{CLMPR18}.
None of our countermeasures or algorithmic improvements 
were applied.

\paragraph{Meyer--Campos--Reith.}
We used the software library freely available from the authors of~\cite{MCR18}.
This software batches isogenies using SIMBA-5-11.
The improvements we describe in~\S\ref{sec:repair} and~\S\ref{sec:optimize}
were \emph{not} applied.

\paragraph{Onuki \emph{et. al.}}
Unfortunately, the source code for the implementation in~\cite{OAYT19}
was not freely available,
so direct comparison with our implementation was impossible.
Table~\ref{tab:ops-revised}
includes their field operation counts
for their unmodified algorithm using SIMBA-3-8.
We did not apply the optimizations of~\S\ref{sec:optimize} here.
(We do not replicate the cycle counts from~\cite{OAYT19}
in Table~\ref{tab:cc-revised},
since they may have been obtained using turbo boost,
thus rendering any comparison invalid.)

\paragraph{Our implementations.}
We implemented three constant-time CSIDH algorithms,
using the standard primes
with the exponent bounds \(m_i\) from~\cite[\S5.2]{OAYT19}.
\begin{description}
    \item[MCR-style]
        This is essentially our version of Meyer--Campos--Reith
        (with one torsion point and dummy operations,
        batching isogenies with SIMBA-5-11),
        but
        applying the techniques of~\S\ref{sec:repair} and~\S\ref{sec:optimize}.
    \item[OAYT-style]
        This is essentially our version of Onuki \emph{et. al.}
        (using two torsion points and dummy operations,
        batching isogenies with SIMBA-3-8),
        but
        applying the techniques of~\S\ref{sec:repair} and~\S\ref{sec:optimize}.
    \item[No-dummy]
        This is Algorithm~\ref{alg:DummyFree}
        (with two torsion points and no dummy operations),
        batching isogenies using SIMBA-5-11.
\end{description}
In each case, the improvements and optimizations of
\S\ref{sec:repair}-\ref{sec:optimize} are applied,
including projective Elligator,
short differential addition chains,
and twisted Edwards arithmetic and isogenies.
Our software library is freely available from 
\begin{center}
    \url{https://github.com/JJChiDguez/csidh}\,. 
\end{center}
The field arithmetic is based on the Meyer--Campos--Reith software library~\cite{MCR18};
since the underlying arithmetic is essentially identical,
the performance comparisons below
reflect differences in the CSIDH algorithms.

\paragraph{Results.}

We see in Table~\ref{tab:cc-revised}
that the techniques we introduced in~\S\ref{sec:repair}
and~\S\ref{sec:optimize}
produce substantial savings compared with the implementation
of~\cite{MCR18}.
In particular,
our OAYT-style implementation yields a 39\% improvement
over~\cite{MCR18}.
Since the implementations use the same underlying field arithmetic library,
these improvements are entirely due to the techniques introduced in this paper.
While our no-dummy variant is (unsurprisingly) slower,
we see that the performance penalty is not prohibitive:
it is less than twice as slow as our fastest dummy-operation algorithm,
and only 22\% slower than~\cite{MCR18}.

\begin{table}[ht]
    \caption{%
        Field operation counts for constant-time CSIDH.
        Counts are given in millions of operations, averaged over 1024 random experiments.
        The performance ratio uses~\cite{MCR18} as a baseline,
        considers only multiplication and squaring
        operations, and assumes $M = S$.
    }
    \label{tab:ops-revised}
    \centering
    \begin{tabular}{c|c@{\;}||@{\;}r@{\;}|@{\;}r@{\;}|@{\;}r@{\;}|@{\;}r}
        \textbf{Implementation} & \textbf{CSIDH Algorithm} & \textbf{M}
        & \textbf{S} & \textbf{A} & \textbf{Ratio}
        \\ \hline
        Castryck et al.~\cite{CLMPR18}
        	& unprotected, unmodified
        	& 0.252 & 0.130 & 0.348 & 0.26
        \\ \hline
        Meyer--Campos--Reith~\cite{MCR18} 
            & unmodified 
            & 1.054 &  0.410 &  1.053 & 1.00
        \\ \hline
        Onuki et al.~\cite{OAYT19}
            & unmodified 
            & 0.733 & 0.244 & 0.681 & 0.67
        \\ \hline
        \multirow{3}{*}{This work} 
            & MCR-style
            & 0.901 & 0.309 & 0.965 & 0.83
        \\ \cline{2-6}
            & OAYT-style
            &  0.657     &  0.210     &  0.691     &        0.59
        \\ \cline{2-6}
            & No-dummy
            & 1.319      &  0.423     &  1.389     &        1.19
        \\ \hline
    \end{tabular}
\end{table} 

\begin{table}[ht]
    \caption{%
        Clock cycle counts for constant-time CSIDH implementations,
        averaged over 1024 experiments. 
        The ratio is computed using~\cite{MCR18} as baseline implementation.
    }
    \label{tab:cc-revised}
    \centering
    \begin{tabular}{c@{\;}|@{\;}c@{\;}|r|r}
        \textbf{Implementation} & \textbf{CSIDH algorithm} & \textbf{Mcycles} & \textbf{Ratio} 
        \\ \hline
        Castryck et al.~\cite{CLMPR18}
        & unprotected, unmodified
        & 155
        & 0.39
        \\ \hline
        Meyer--Campos--Reith~\cite{MCR18} 
        & unmodified 
        & 395 
        & 1.00
        \\ \hline
        \multirow{3}{*}{This work} 
        & MCR-style 
        & 337
        & 0.85
        \\ \cline{2-4}
        & OAYT-style
        & 239
        & 0.61
        \\ \cline{2-4}
        & No-dummy
        & 481
        & 1.22
        \\ \hline
    \end{tabular}
\end{table}

\section{
    Conclusion and perspectives
}
\label{sec:conclusion}

We studied side-channel protected implementations of the isogeny based
primitive CSIDH. %
Previous implementations failed at being constant time because of a
subtle mistake. %
We fixed the problem, and proposed new improvements, to achieve the
most efficient version of CSIDH protected against timing and simple power
analysis attacks to date.
All of our algorithms were implemented in C, and the source made publicly
available online.

We also studied the security of CSIDH in stronger attack scenarios. %
We proposed a protection against some fault-injection and timing attacks
that only comes at a cost of a twofold slowdown. %
We also sketched an alternative version of CSIDH ``for the paranoid'',
with much stronger security guarantees, however at the moment this
version seems too costly for the security benefits; more work is
required to make it competitive with the original definition of CSIDH.

\bibliography{draft_CT}
\bibliographystyle{plain}
\end{document}